\documentclass[final,usletter,number,sort&compress]{elsarticle}

\usepackage{graphicx}
\usepackage{dcolumn}
\usepackage{bbm}
\usepackage{textcomp}
\usepackage{color}
\usepackage{gensymb}
\usepackage{amsmath}
\usepackage{wasysym}

\begin{document}

\def\nuc#1#2{${}^{#1}$#2}
\def\mc{MiniCLEAN}

\title{Fluorescence Efficiency and Visible Re-emission Spectrum of Tetraphenyl Butadiene Films at Extreme Ultraviolet Wavelengths}
\author[lanl]{V. M. Gehman\corref{corvmg}}
\author[lanl,penn]{S. R. Seibert\corref{corsrs}}
\author[lanl]{K. Rielage}
\author[lanl]{A. Hime}
\author[usd]{Y. Sun}
\author[usd]{D.-M. Mei}
\author[dsu]{J. Maassen}
\author[dsu]{D. Moore}

\address[lanl]{Los Alamos National Laboratory, Los Alamos, NM 87545}
\address[penn]{University of Pennsylvania, Philadelphia, PA 19104}
\address[usd]{University of South Dakota, Vermillion  SD  57069}
\address[dsu]{Dakota State University, Madison, SD 57042}
\cortext[corvmg]{Corresponding author: vmg@lanl.gov}
\cortext[corsrs]{Corresponding author: sseibert@hep.upenn.edu}
\date{\today}

\begin{abstract}
A large number of current and future experiments in neutrino and dark matter detection use the scintillation light from noble elements as a mechanism for measuring energy deposition.  The scintillation light from these elements is produced in the extreme ultraviolet (EUV) range, from 60--200 nm.  Currently, the most practical technique for observing light at these wavelengths is to surround the scintillation volume with a thin film of Tetraphenyl Butadiene (TPB) to act as a fluor.  The TPB film absorbs EUV photons and reemits visible photons, detectable with a variety of commercial photosensors.  Here we present a measurement of the re-emission spectrum of TPB films when illuminated with 128, 160, 175, and 250 nm light.  We also measure the fluorescence efficiency as a function of incident wavelength from 120 to 250 nm.
\end{abstract}


\maketitle

\section{Introduction and Motivation}\label{sec:Intro}
Detecting the scintillation light from liquid noble elements (``noble liquids'') is an important component of several large experimental programs engaged in efforts to directly detect dark matter\cite{warp,ardm,miniclean} as well as neutrinos from both the Sun\cite{clean} and accelerator beam-lines\cite{icarus,microb,lbne}.  Liquid noble elements have a number of characteristics that make their use as scintillators very attractive.  They have good stopping power compared to other liquid scintillators, with densities between 1.2 and 3.1 g/cm$^{3}$\cite{noblegas}.  Their scintillation yield is also very high, typically tens of photons per keV\cite{neonscint} (similar to that of NaI\cite{Knoll}).  The time structure of scintillation light from the noble elements also offers excellent particle identification capability, a property that will be discussed in Section \ref{sec:NEScint}.  Lastly, noble liquids are essentially transparent to their own scintillation light, eliminating the need to dissolve fluors directly into the scintillation volume, as is the case with most organic liquid scintillators\cite{birks1953}.

The primary difficulty in using noble elements as a scintillator comes not from the cryogenics required to keep them in a liquid state, but from the challenges of efficiently detecting the short wavelength scintillation light.  Noble elements scintillate in the extreme ultraviolet (EUV) wavelength range.  Most current interest in the community involves xenon and argon, which have scintillation emission spectra ranging from roughly 120 to 200 nm.  As such, the bulk of the work presented in this article will focus on this wavelength band.  There is also interest in using neon\cite{clean} and helium\cite{nEDM,Golub94} as scintillators, both of which emit scintillation light at much shorter wavelengths, from 60 to 100 nm.  One other difficulty associated with the use of noble elements as scintillators is purification.  Very small quantities of nitrogen, oxygen or water can strongly quench scintillation light\cite{Acciarri2009}.  Purification is largely a solved problem in terms of methodology.  Implementation of the necessary purification regime can however be somewhat technically challenging and should not be ignored.

The detection of EUV photons is a very challenging problem.  Their wavelength is short enough that they are strongly absorbed by nearly all materials used for visible optics, such as the quartz or glass windows of photomultiplier tubes, but they are not energetic enough to be treated calorimetrically like x rays or $\gamma$ rays.  This obstacle can be side-stepped by down-scattering EUV photons into the visible wavelength band with a fluor.  One very common technique is to coat a surface in contact with the noble element with a thin film of the organic compound tetra-phenyl butadiene (TPB).  This article will explore the efficiency with which a TPB film converts EUV to visible photons as well as the spectrum of the re-emitted photons as a function of the wavelength of the absorbed EUV photons.

We will begin Section \ref{sec:ThePast} with a discussion of noble element scintillation.  We will present a summary of previous measurements involving TPB films, including both the fluorescence efficiency and the visible re-emission spectrum.  We will also discuss how previous noble element scintillation detectors avoided the problem of the large uncertainty on both of these quantities.  Section \ref{sec:Hardware} will discuss the experimental apparatus used for this study, with special attention paid to the improvements made over previous measurements.  Section \ref{sec:VisSpec} will detail our measurements of the TPB visible re-emission spectrum as a function of input wavelength, and Section \ref{sec:FlEff} will present our fluorescence efficiency measurements.  Last, we will draw some conclusions from this work and discuss our plans for future measurements in Section \ref{sec:Concl}.

\section{Previous Measurements}\label{sec:ThePast}

\subsection{Noble Element Scintillation}\label{sec:NEScint}
\begin{figure}[htbp]
\begin{center}
\includegraphics[angle=0, width=\textwidth]{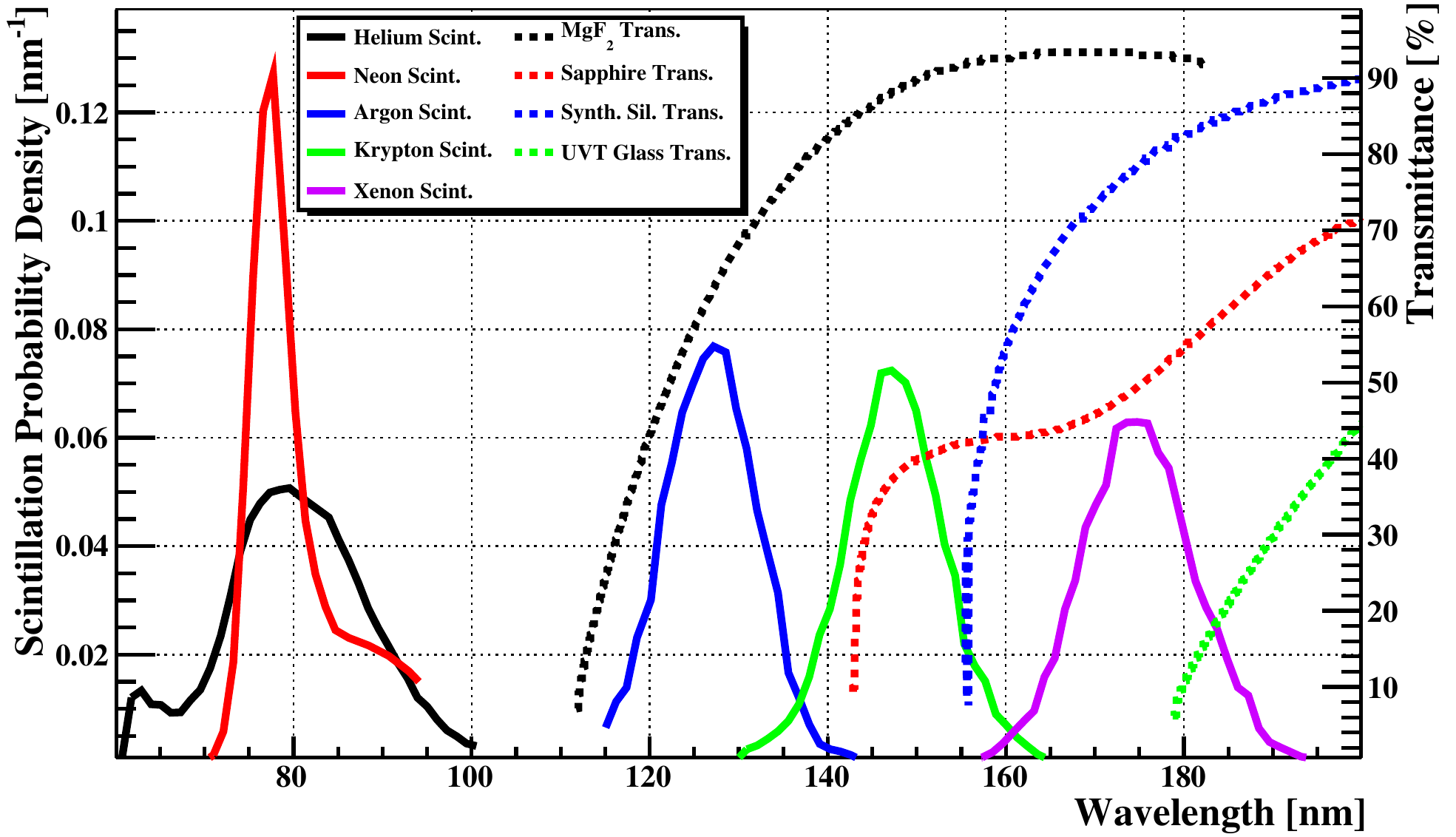}
\caption{Left axis/solid lines: scintillation spectra for helium (reproduced from \cite{Stockton1970}), neon (reproduced from \cite{Packard1970}), argon, krypton and xenon (reproduced from \cite{Saito2002}).  Right axis/dotted lines: percent transmittance of several common optical window material (reproduced from \cite[Figure 4.5]{PMTHandbook}).}
\label{fig:ScintSpec}
\end{center}
\end{figure}

As discussed above, noble elements are transparent to their own scintillation light, unlike organic scintillators.  This is due to the somewhat unusual way in which scintillation occurs in noble elements.  This is discussed in detail in \cite{Lippincott2010}.  Rather than simply exciting an atomic or molecular state which then emits light as it relaxes to the ground state, scintillation in noble elements is actually produced by the de-excitation of dimer molecules formed from combinations of excited or ionized atoms and a ground state atom.  This scintillation light has two components arising from two different states: a short-lived, singlet state, and a long-lived, triplet state.  The lifetime difference comes from whether the noble atom in question is ionized or simply excited when it forms the dimer molecule.  Dimers formed from an excited atom can decay very quickly with a typical lifetime of $\apprle 10$ ns.  Ionized dimers, on the other hand, must overcome a spin flip before the dimer state can decay.  This dramatically increases the lifetime of the dimer to microseconds or longer.  Because the scintillation light corresponds to a transition in a temporary dimer state, there is no corresponding transition in the surrounding monoatomic medium to absorb photons once produced.  As a result, these scintillation photons have a very long path length in the bulk liquid, making noble elements essentially transparent at their scintillation wavelengths.  Figure \ref{fig:ScintSpec} shows the scintillation spectra of all five noble elements along with the percent transmittance of several common optical window materials.  The data presented in Figure \ref{fig:ScintSpec} are reproduced from References \cite{Stockton1970, Packard1970, Saito2002, PMTHandbook}.  Each spectrum has been normalized to unit area to facilitate comparison of their shapes.

\subsection{TPB Re-emission Spectrum}\label{sec:PrevVisSpec}
Several measurements of the visible re-emission spectrum of TPB films have been made\cite{Lally1996, Burton1973}.  However, the TPB was excited with UV sources at 253.7 nm \cite{Burton1973} and 185 nm \cite{Lally1996}, which are both longer wavelength than the scintillation spectrum of argon, neon, and helium.  Furthermore, the spectra measured in these references are quite different, leading to the possibility of a significant dependence on the input EUV wavelength or the details of the TPB deposition method.  Rather than use one of these existing measurements of the re-emission spectrum as an input to the fluorescence efficiency measurement, we directly measured the re-emission spectrum of our TPB samples using several excitation wavelengths from 128 nm to 250 nm.

\subsection{TPB Fluorescence Efficiency}\label{sec:PrevFlEff}
There have also been several measurements of the fluorescence efficiency of TPB films.  Reference \cite{Burton1973} measured this efficiency relative to sodium salicylate, for several TPB coating thicknesses, illuminating the films with EUV light from about 100 to above 300 nm.  To convert this to an absolute efficiency, one requires the absolute efficiency of sodium salicylate.  Reference \cite{Burton1973} refers its readers to Reference \cite[Table 7.1]{Samson} for these efficiency numbers.  Unfortunately, the absolute efficiency numbers in \cite[Table 7.1]{Samson} differ by more than a factor of three.  Reference \cite{Regan1994} also measures the fluorescence efficiency of TPB along with several other fluors, but does so at much shorter wavelengths, ranging from 0.989 to 6.76 nm.  The uncertainties of the efficiencies measured in Reference \cite{Regan1994} are quite good compared to most others, at roughly 25\%.  TPB represents a significant improvement over other fluors (like sodium salicylate) because it can be applied to a substrate via vacuum deposition rather than by wet-dipping a slide.  This dramatically enhances film reproducibility as allowing for cleaner application of the wavelength shifter.  This is particularly important in low-background applications like direct dark matter detection, where the low levels of naturally occurring uranium and thorium can serve to obscure the signal of interest.

\section{Experimental Apparatus}\label{sec:Hardware}
\begin{figure}[htbp]
\begin{center}
\includegraphics[angle=0, width=\textwidth]{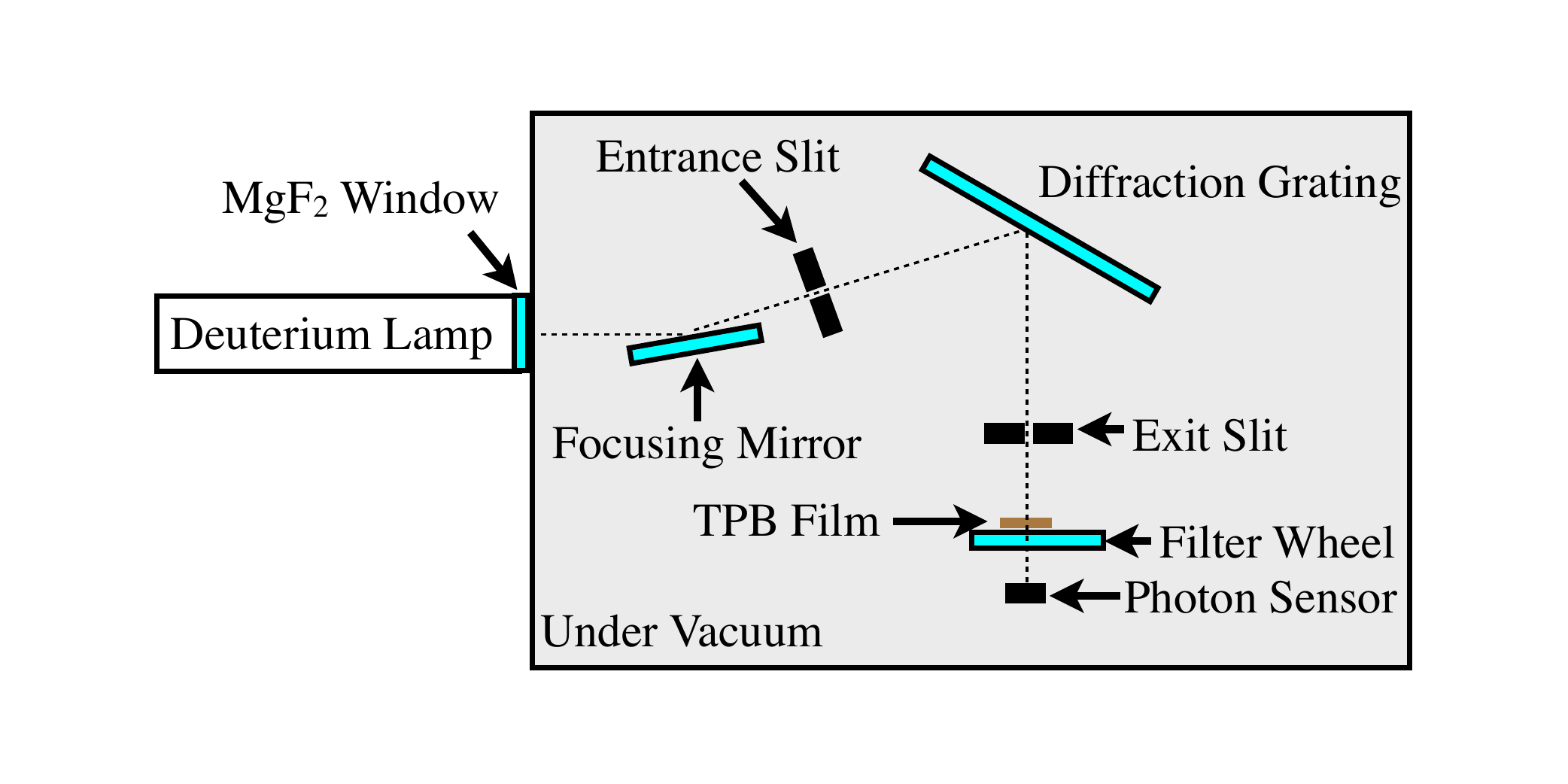}
\caption{Diagram of the experimental apparatus.  The parts of apparatus shaded gray are under vacuum.  The deuterium arc lamp is separated from the rest of the light source by a MgF$_{2}$ window, and the TPB film was evaporated onto the side of the sample disc facing the monochromator.}
\label{fig:ExpApp}
\end{center}
\end{figure}

Our experimental apparatus consists of three stages:  an EUV light source, a filter wheel with several 2.5 cm diameter acrylic sample disks, and then one of two photon sensors.  A diagram of the experimental apparatus can be found in Figure \ref{fig:ExpApp}.
In the first configuration, light was observed by a UV/visible spectrometer, and in the second configuration, a calibrated, silicon photodiode cell was installed.  In order to minimize the effect of attenuation on our sensitivity, the entire space from the exit window of the deuterium lamp to the photon sensor was kept at a very low pressure, never exceeding $2.5 \times 10^{-6}$ mBar.  The vacuum space also included the entire filter wheel assembly, and the photon sensor.  Additionally, in the first configuration, the spectrometer was coupled to the vacuum space through a collimating lens and fiber feedthrough.

\subsection{Light Source}
\begin{figure}[htbp]
\begin{center}
\includegraphics[angle=0, width=\textwidth]{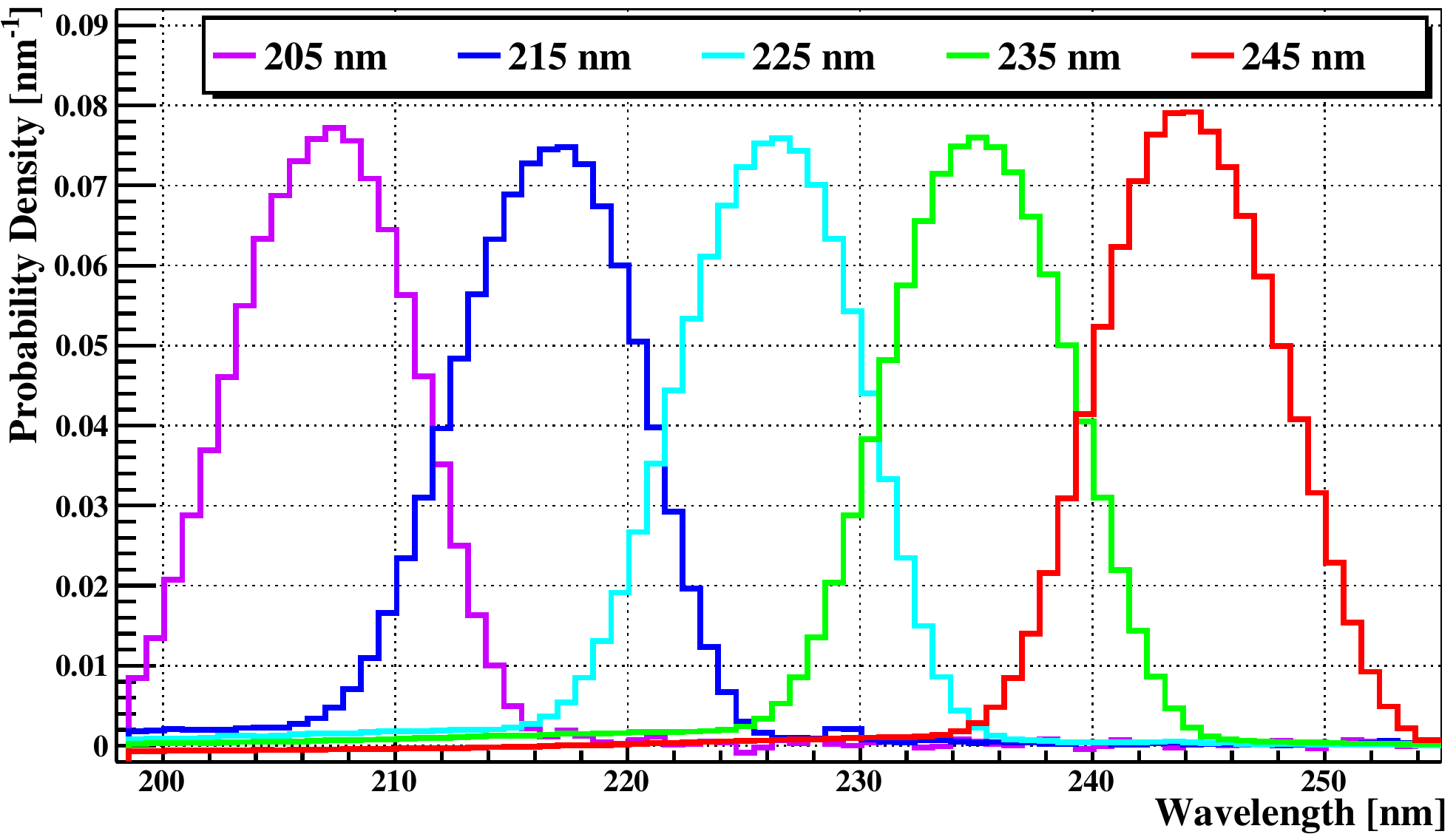}
\caption{Direct EUV light spectra from the monochromator with a 3-mm wide exit aperture.  The wavelength distribution at each setting is approximately Gaussian.}
\label{fig:DirectLamp}
\end{center}
\end{figure}

The light source used in our measurement was a Model 632 Deuterium Light source, combined with a Model 234/302VM 0.2-Meter EUV Monochromator from McPherson, Inc.  This light source has strong peaks at 128 nm and 160 nm, with a long high-wavelength tail that extends up to approximately 250 nm.  The MgF$_{2}$ window separating the deuterium arc lamp from the rest of the light source means that the intensity of the lamp cuts off at 110 nm (see Figure \ref{fig:ScintSpec} for the transmittance of MgF$_{2}$ as a function of wavelength).  The monochromator was built around a 1200 G/mm holographic diffraction grating, which is rotated with respect to the instrument's entrance and exit slits to select specific wavelengths.  Both slits are 10 mm high, and have adjustable widths from 0.01 mm to 3 mm.  We took data for all measurements with the entrance and exit slits set to 3 mm to maximize the light output of our system.  There is a range from 205 nm to 245 nm where our visible spectrometer is sensitive to the direct light from the lamp (see Figure \ref{fig:DirectLamp}).  In this range, we measure the full-width, half-maximum resolution of the monochromator to be $8.5 \pm 0.5$ nm for our slit configuration, after subtracting the resolution of the spectrometer itself.  We use this for our estimate of the monochromator resolution at all wavelengths.

\subsection{Filter Wheel and TPB Samples}\label{sec:FilterWheel}
After leaving the exit slit of the light source, the monochromatic light entered a Model 648 Vacuum Filter Wheel, also from McPherson.  This allowed for selection between an open port, a fully absorbing metal shutter or one of several TPB-coated acrylic disks.  In the open port configuration, the light source intensity could be observed directly by the installed photosensor.  The metal shutter allowed a dark reading to be made with each photosensor and subtracted.  The ability to make both of these measurements without breaking the vacuum seal was critical to minimizing systematic uncertainties from time variation of the light source and the photosensors.  The light source was observed to drop in intensity by 3.6\% per hour during data collection.  As a result, the intensity of the lamp was always measured with the filter wheel in the open position no more than a few minutes before the measurement of the re-emitted light with a TPB disk in position.  The residual time variation is taken as an uncertainty.

\begin{figure}[htbp]
\begin{center}
\includegraphics[angle=0, width=\textwidth]{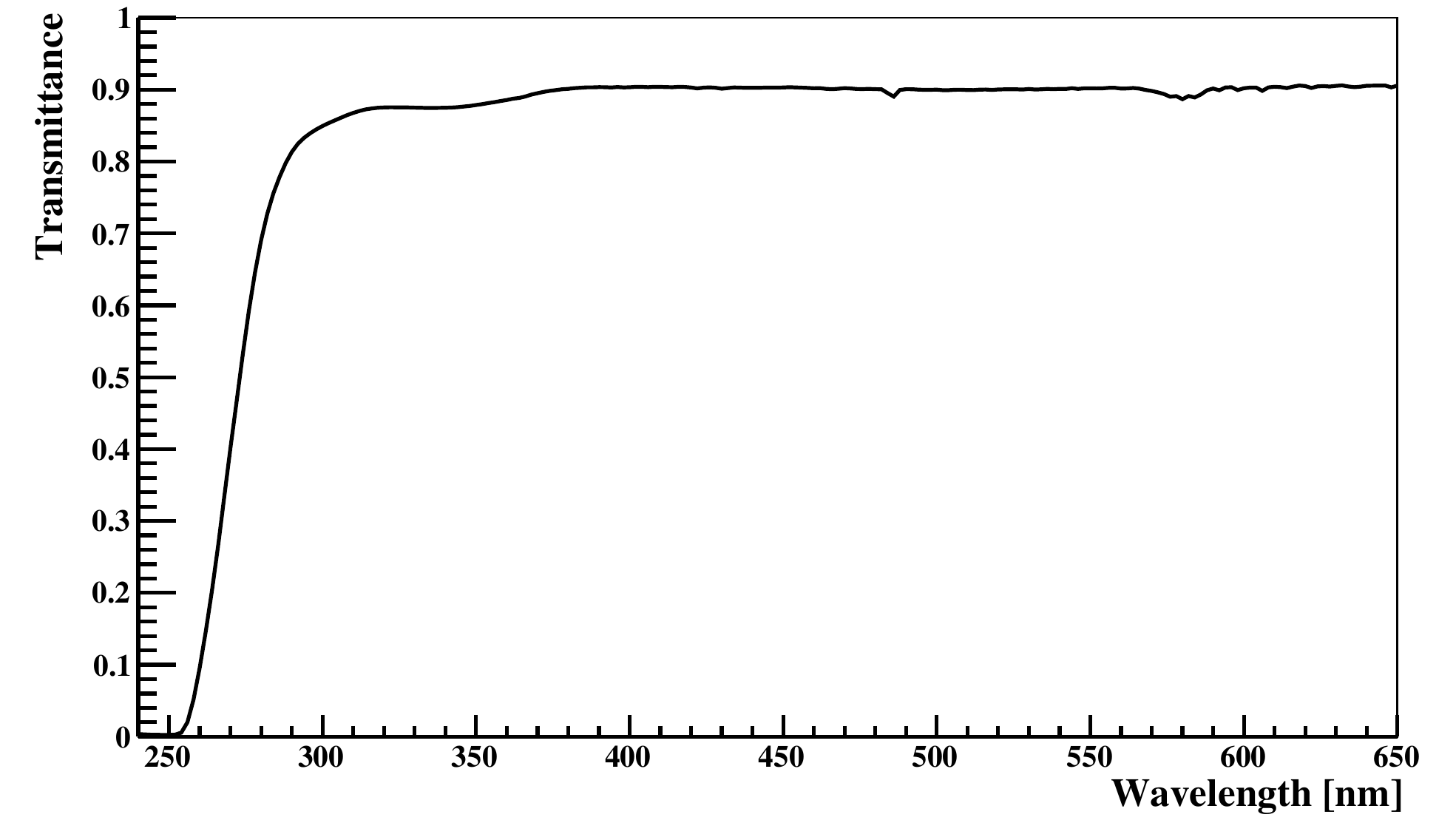}
\caption{Transmittance through 0.47 cm of Solacryl SUVT acrylic in air as a function of wavelength.}
\label{fig:TransmitSUVT}
\end{center}
\end{figure}

The sample TPB film was evaporated onto the face of the acrylic disc facing the light source, because the acrylic is itself opaque to EUV (see Figure \ref{fig:TransmitSUVT}).  The TPB film was deposited on circular disks 2.5 cm in diameter and 0.47 cm in thickness.  To allow a wide range of re-emission wavelengths to pass through the disk to the photosensors, the disks were made of Solacryl SUVT acrylic, manufactured by Spartech Polycast.  We measured the transmittance to be greater than 80\% for wavelengths longer than 290 nm, as shown in Figure \ref{fig:TransmitSUVT}.  The index of refraction of the acrylic is listed by the manufacturer's data sheet to be 1.49\cite{SolacrylDatasheet}.

The TPB was scintillation grade, purchased from Aldrich, and the film was created on the acrylic disks via vacuum deposition.  For this process, TPB powder was placed in a metal boat inside a vacuum bell jar, which then pumped down to a pressure of $5 \times 10^{-5}$ mBar.  The tungsten chimney boat was then heated for approximately fifteen to twenty minutes, evaporating TPB onto the acrylic sample to a film thickness of 0.22 mg/cm$^{2}$.  This film thickness was measured directly to be $1.5 \pm 0.05$ $\mu$m using reflectometry.  The film thickness was chosen to maximize light yield after a series of optimization measurements done with prototype detectors and test benches.  The film tested in this work here is much thinner than previous studies outlined in \cite{Samson}.  All tests described in this article were performed on the same sample.  It is important to emphasize the fact that TPB films prepared via vacuum deposition will behave rather differently than those prepared via other methods (such as wet dipping or some kind of process involving a solvent or paint).  While films prepared via vacuum deposition are generally viewed as more uniform and repeatable than those prepared via some other methods, no attempt was made to check the film thickness for variability across the disk.  The reflectometry measurement was however performed at the same spot on the disc illuminated by the exit slit.  It is also true that serial testing of multiple TPB films, prepared under ostensibly identical circumstances would aid in understanding ``film to film'' variability.

\subsection{Spectrometer}
\begin{figure}[htbp]
\begin{center}
\includegraphics[angle=0, width=\textwidth]{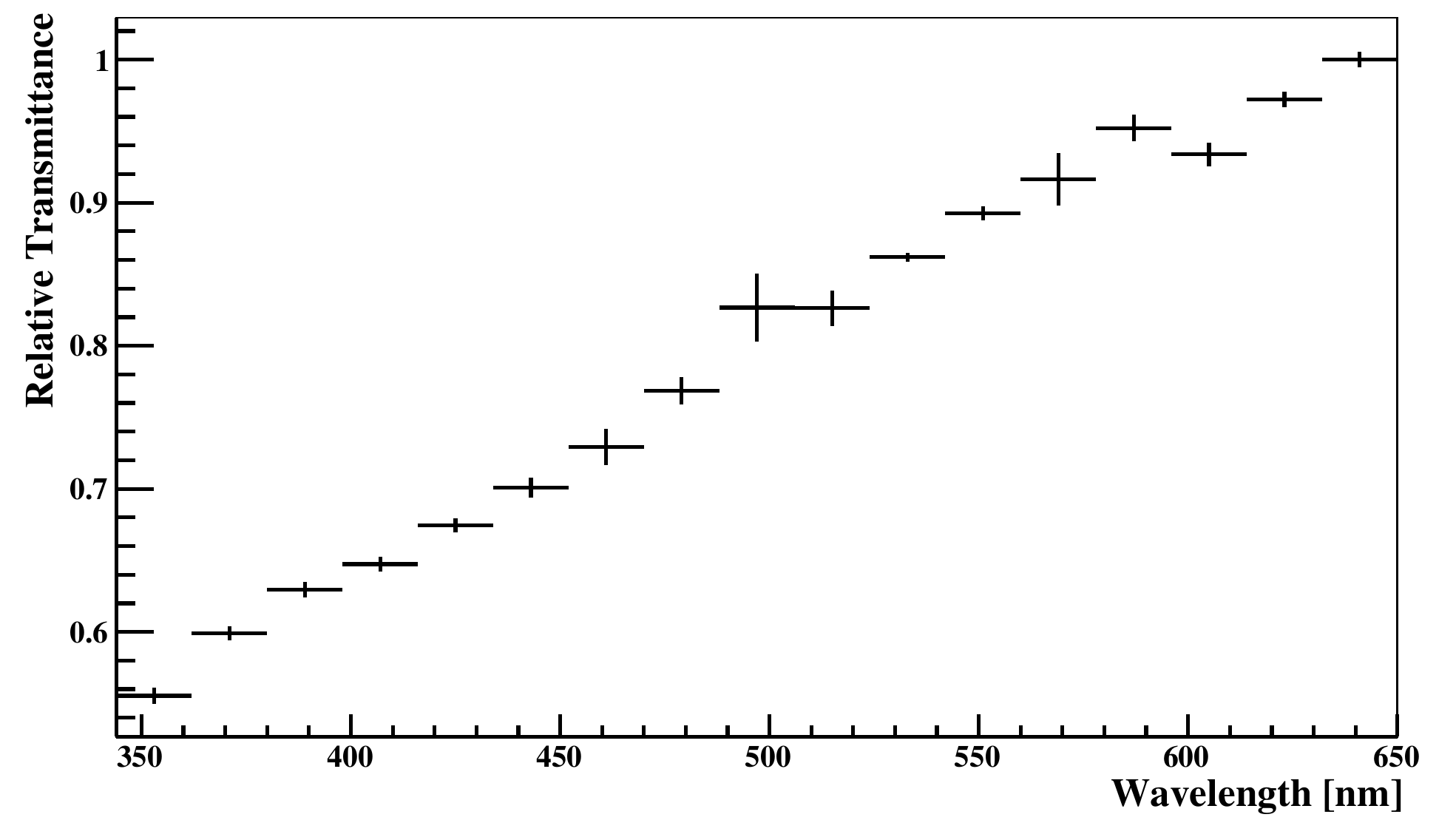}
\caption{Relative transmittance of collimating lens, fiber optic vacuum feedthrough, and 200 $\mu$m fiber leading to the spectrometer.  The function has been normalized so that the highest value is 1.}
\label{fig:SpecOpticsTrans}
\end{center}
\end{figure}

The visible re-emission spectra were taken with a QE65000 spectrometer from Ocean Optics.  In order to direct light from the vacuum chamber containing the light source and TPB samples, a collimating lens was used to focus light onto a fiber optic vacuum feedthrough out to a 200 $\mu$m diameter fiber leading to the spectrometer input port.  Spectra were dark-subtracted in the data acquisition software and analyzed with the ROOT analysis framework\cite{root}. A broadband xenon lamp was used to calibrate the relative transmittance of the optical train leading to the spectrometer, shown in Figure \ref{fig:SpecOpticsTrans}.  The xenon lamp was observed through a fiber with a transmittance function calibrated by the manufacturer.  Then the spectrum of the same lamp was observed through the complete optical train, divided by the raw lamp spectrum, and rescaled to produce the relative transmittance function.

\subsection{Photodiode}
\begin{figure}[htbp]
\begin{center}
\includegraphics[angle=0, width=\textwidth]{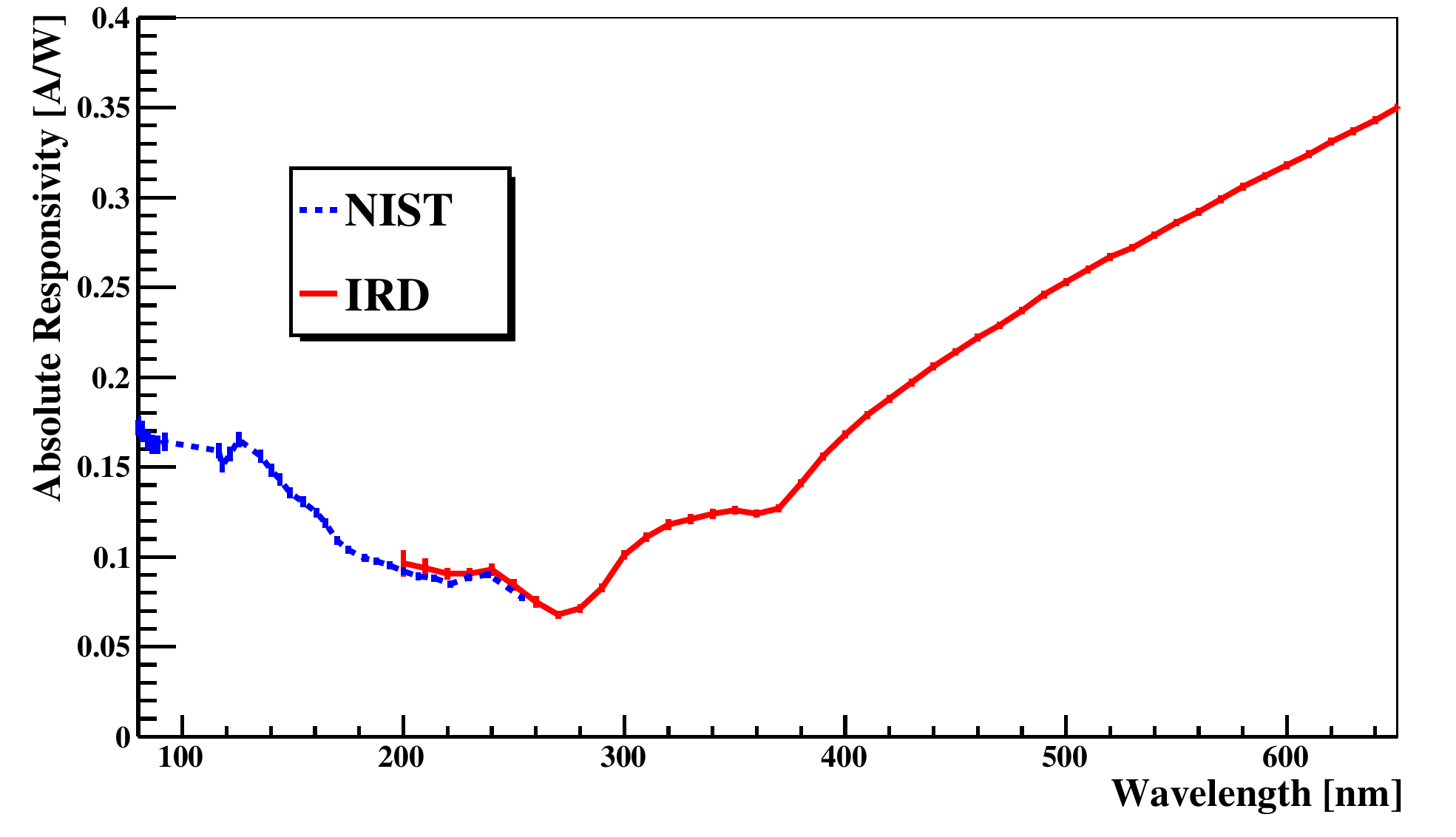}
\caption{Calibrated current response of AXUV100G photodiode as a function of incident wavelength as measured by NIST and IRD.}
\label{fig:AXUVCalibration}
\end{center}
\end{figure}

The absolute intensity measurements were made with an AXUV100G photodiode from International Radiation Detectors, Inc (IRD)\cite{Canfield1998, Gillikson1996}.  This device has a one centimeter by one centimeter active photodiode area.  It is a windowless silicon photodiode cell, with a calibrated sensitivity from 80 nm to 1100 nm.  The device is completely passive, so it was read out with a Keithley 6487 picoammeter.  The calibration, provided by NIST for incident light from 80 to 254 nm and IRD for light from 200 to 1100 nm, scales a measured current into a radiant flux as a function of wavelength.  Figure \ref{fig:AXUVCalibration} shows the wavelength dependence of the photodiode response.

\section{Visible Re-emission Spectrum}\label{sec:VisSpec}
\begin{figure}[htbp]
\begin{center}
\includegraphics[angle=0, width=\textwidth]{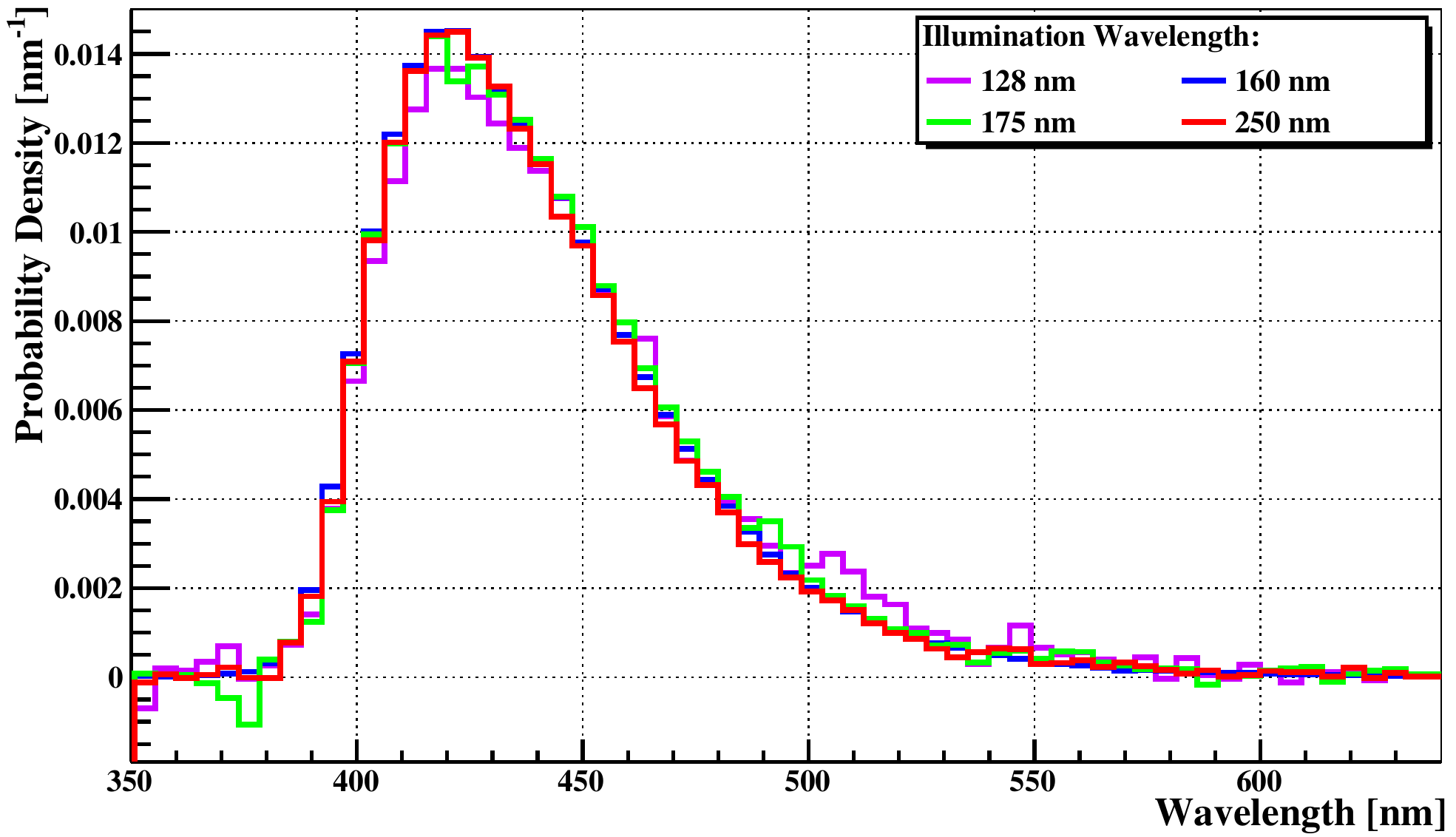}
\caption{Visible re-emission spectrum for a TPB film illuminated with 128, 160, 175, and 250 nm light.  All spectra are normalized to unit area.}
\label{fig:VisSpec}
\end{center}
\end{figure}

We captured visible re-emission spectra for input EUV wavelengths of: 128, 160, 175, and 250 nm.  These wavelengths correspond to: the peak of the argon scintillation spectrum, a bright peak in the emission spectrum of our light source, the peak of the xenon scintillation spectrum, and the peak of the emission spectrum of most ultraviolet light emitting diodes.  The normalized visible re-emission spectra captured for all four input wavelengths are presented in Figure \ref{fig:VisSpec}.

Upon examination of Figure \ref{fig:VisSpec}, we find no strong incident wavelength dependence in the shape of the visible re-emission spectra.  All four spectra show a strong cut-off at short wavelength at roughly 400 nm.  The only slight difference appears in the 128 nm spectrum, which shows a slight excess above the others at just over 500 nm.  We use the 128 nm re-emission spectra to calculate the fluorescence efficiency for all incident wavelengths in Section \ref{sec:FlEff}.

\section{Fluorescence Efficiency}\label{sec:FlEff}
We define the fluorescence efficiency of the TPB layer as a function of incident wavelength, $\epsilon(\lambda)$, to be the ratio of the number of re-emitted visible photons to the number of UV photons incident on the TPB layer.  The efficiency at a given wavelength, $\lambda$, is computed from the current produced by the photodiode when directly viewing the light source, $I_{\rm lamp}(\lambda)$, the current when viewing the opaque shutter, $I_{\rm dark}(\lambda)$, and the current when viewing the acrylic disk with TPB film on it, $I_{\rm TPB}(\lambda)$.  The dark-subtracted current ratio is converted to an efficiency using the equation
\begin{equation}\label{eq:efficiency}
\epsilon(\lambda) = \frac{ I_{\rm TPB} - I_{\rm dark}}{I_{\rm lamp} - I_{\rm dark}} \times g\frac{ \int d\lambda' \frac{hc}{\lambda'} C(\lambda') S(\lambda - \lambda')}{  \int d\lambda'' \frac{hc}{\lambda''} C(\lambda'') R(\lambda'')},
\end{equation}
where $g$ is a geometric efficiency based on the configuration of the apparatus, $C(\lambda)$ is the photodiode response function shown in Figure \ref{fig:AXUVCalibration}, $S(\lambda - \lambda')$ is the Gaussian wavelength distribution of the monochromator centered around $\lambda$, and $R(\lambda'')$ is the re-emission spectrum of TPB shown in Figure \ref{fig:VisSpec}.

The geometric efficiency, $g$, is the ratio of the number of UV photons reaching the photodiode when the filter wheel is in the open position to the number of visible photons observed by the photodiode when a TPB sample is in position, assuming unit fluorescence efficiency.  This constant is independent of the wavelength of the incident EUV light and the re-emission spectrum of visible light.  In general, $g$ is different from 1 due to the different angular distributions of incident and reemitted light, the effects of refraction at the acrylic-vacuum interface of the TPB sample disk, and the limited solid angle of the photodiode.  We evaluate $g$ using a simple photon-tracing Monte Carlo simulation that initiates photons at the diffraction grating of the monochromator, propagates them through the exit aperture to the 2.3 cm diameter hole in the filter wheel where they can interact with a TPB layer deposited on an acrylic disk before arriving at the photodiode.  Fresnel reflection and refraction at the TPB and acrylic surfaces is included in the Monte Carlo, but reflection at the various interior black walls of the apparatus are neglected as second-order effects.

Not only does the geometric efficiency depend on the physical location and size of objects in the measurement apparatus, but it also depends on the intrinsic angular distribution of re-emitted light from the TPB film.  The angular distribution of re-emission from vapor-deposited thin films of TPB has not been measured yet, so we would like to report an efficiency measure that is independent of this unknown angular distribution.  Therefore, we define the forward efficiency, $\mathcal{F}(\lambda)$, to be 
\begin{equation}\label{eq:forward_efficiency}
 \mathcal{F}(\lambda) =  \frac{ I_{\rm TPB} - I_{\rm dark}}{I_{\rm lamp} - I_{\rm dark}} \times g'\frac{ \int d\lambda' \frac{hc}{\lambda'} C(\lambda') S(\lambda - \lambda')}{  \int d\lambda'' \frac{hc}{\lambda''} C(\lambda'') R(\lambda'')},
\end{equation}
where $g'$ is the ratio of the number of UV photons reaching the photodiode when the filter wheel in the open position to the number of UV which reach the TPB surface.  The arrangement of our apparatus is such that $g'$ is exactly 1.  By defining forward efficiency in this manner, the total efficiency can be calculated given an angular distribution of TPB re-emission,
\begin{equation}
  \epsilon(\lambda) = \frac{\mathcal{F}(\lambda)}{A},
\end{equation}
where $A$ is the acceptance fraction of a 1 cm by 1 cm square illuminated by the TPB angular distribution from a distance of 2.1 cm away after passing through 7 mm of SUVT acrylic.

\begin{figure}[htbp]
\begin{center}
\includegraphics[angle=0, width=\textwidth]{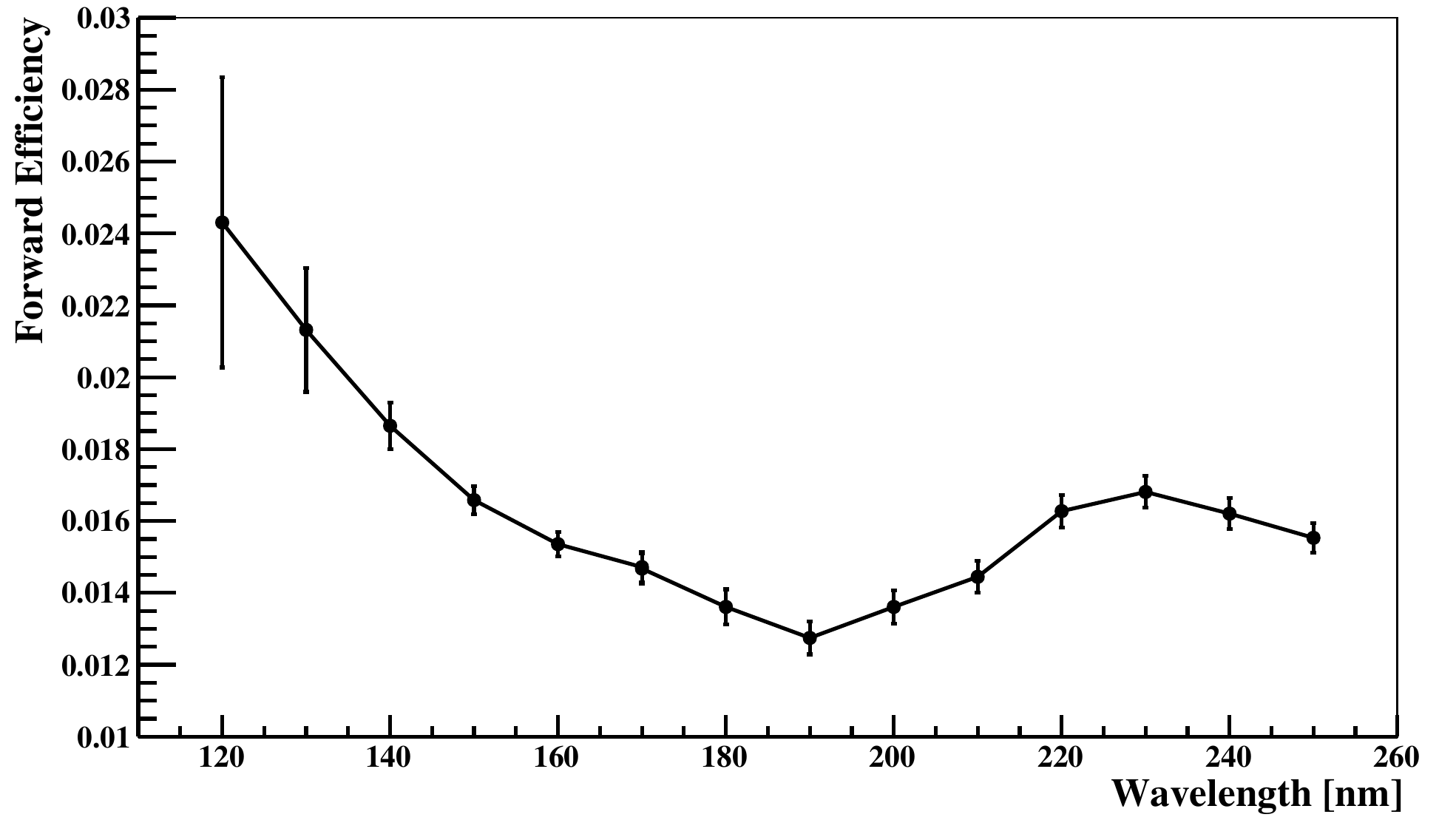}
\caption{\label{fig:ForwardEfficiency}Number of visible photons observed at photodiode sensor per incident EUV photon a function of input EUV photon wavelength.}
\end{center}
\end{figure}

The forward efficiency of our TPB film is presented in Figure \ref{fig:ForwardEfficiency}.  There is a general trend below 190 nm toward increasing fluorescence efficiency with decreasing UV wavelength.  To illustrate the relative sizes of the components of the overall uncertainty, Table \ref{tab:uncert} shows the fractional uncertainties for the TPB efficiency with 130 nm incident light.  The statistical uncertainty in all current measurements was estimated by considering the RMS variation in the current values taken with the metal shutter closed throughout the data collection process.  This uncertainty was propagated through dark-subtracted current ratio from Equation \ref{eq:forward_efficiency}.  The systematic uncertainty caused by short term variation in the lamp was estimated by observing the lamp intensity at 120 nm, where the effect was most significant, at different times and dividing this difference over the time between measurements at the other wavelengths. The final two systematic uncertainties account for the uncertainty in the response of the photodiode integrated over the ultraviolet and visible bands.  This includes not only the uncertainty in the calibration, but also the uncertainty in the spectrum of the light source and TPB re-emission.  By far, the dominant uncertainty in the overall measurement is statistical, caused by relatively large fluctuations in the current measurement relative to the low intensity of the light source at 130 nm.

\begin{table}[htbp]
\begin{center}
\begin{tabular}{|l|r|} \hline
Source of Uncertainty & Uncertainty on $\epsilon$ \\ \hline \hline
Statistical uncertainty in dark-subtracted current ratio  & 7.7\% \\ \hline
Short time variation in the lamp &  0.3\% \\ \hline
Calibration uncertainty of photodiode at 130 nm & 2.3\% \\ \hline
Calibration uncertainty of photodiode at 425 nm & 0.8\% \\ \hline
Total & 8.1\% \\ \hline
\end{tabular}
\caption{\label{tab:uncert}Uncertainties on forward efficiency per unit solid angle at 130 nm.}
\end{center}
\end{table}

\begin{figure}[htbp]
\begin{center}
\includegraphics[angle=0, width=\textwidth]{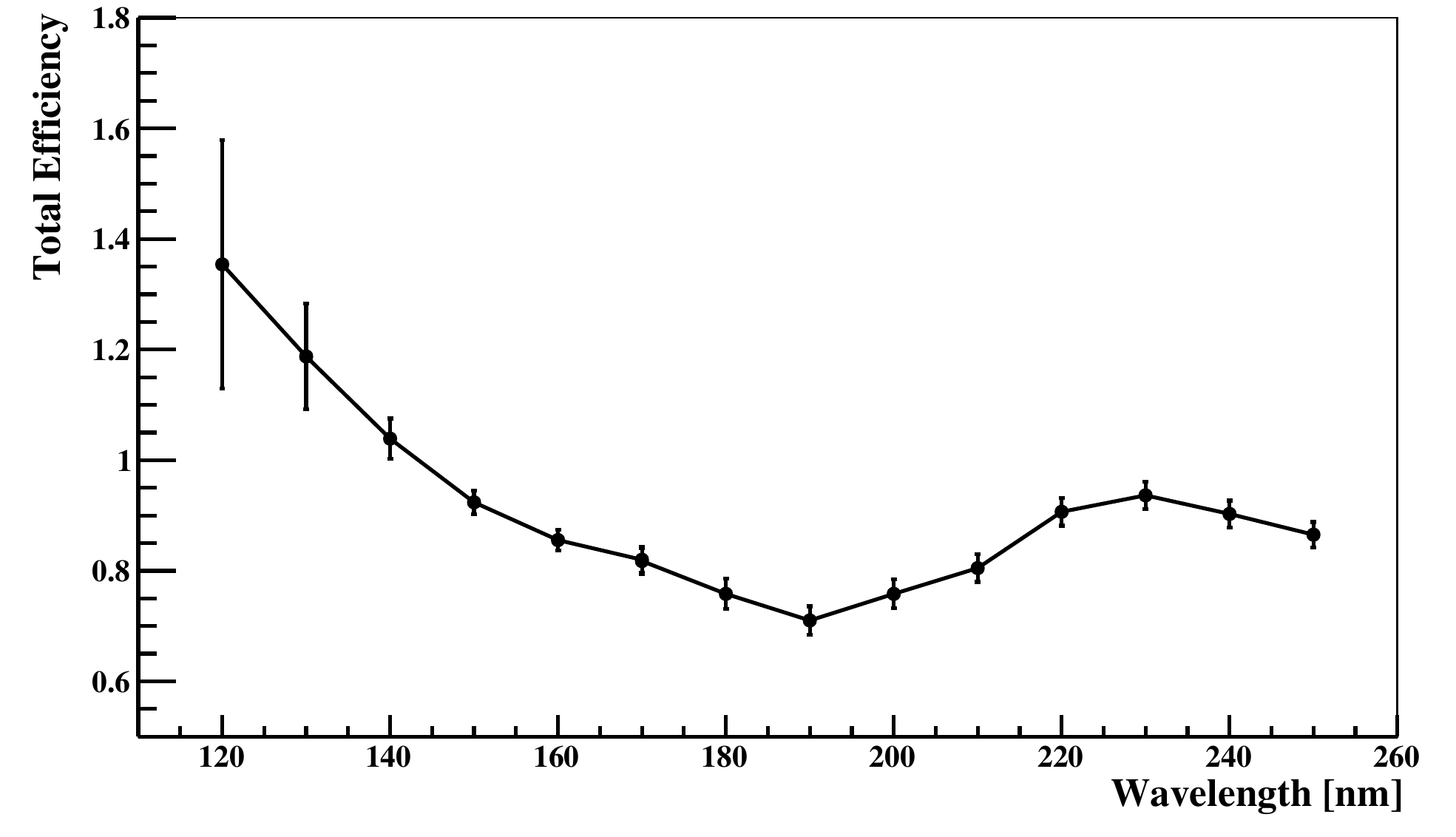}
\caption{Total integrated fluorescence efficiency as a function of input EUV photon wavelength, assuming a Lambertian angular distribution of reemitted light from both sides of the TPB film, similar to sodium salicylate.}
\label{fig:TotalEfficiency}
\end{center}
\end{figure}

To illustrate what the total fluorescence efficiency of TPB could be given a reasonable assumption about the angular distribution of re-emission, we can look to a similar fluor: sodium salicylate.  For film thicknesses that maximize fluorescence efficiency, the angular distribution of re-emitted photons from sodium salicylate follows a cosine law (or Lambertian) distribution in both the forward and backward directions\cite{Allison1964}.  We can apply this same distribution to TPB in order to estimate the total fluorescence efficiency.  Figure \ref{fig:TotalEfficiency} shows the total efficiency of TPB fluorescence computed from our measurement under this assumption.  The acceptance fraction, $A$, is 0.0179 in this case.

Under the assumption of Lambertian re-emission, we find variation an average value of the total fluorescence efficiency between 0.7 and 1.35 visible photons emitted for each EUV photon absorbed.  This reaches a minimum value of approximately 0.7 at 190 nm.  There is a strong upturn in the fluorescence efficiency at short wavelengths, leading to a value of approximately 1.2 at 128 nm.

\section{Conclusions}\label{sec:Concl}
We have presented measurements of both the visible re-emission spectrum and fluorescence efficiency of thin TPB films evaporated onto acrylic.  The visible spectrum we measure here is somewhat closer to that presented in Reference \cite{Lally1996} than it is to the one from \cite{Burton1973}, although our spectrum seems cuts off at a somewhat shorter wavelength than seems apparent in \cite{Lally1996}.

The details of the work presented in this article are somewhat specific to the \mc\ experiment in terms of the preparation of the TPB film.  The next step in TPB characterization is to measure the angular distribution of re-emission in order to test the assumption of a Lambertian distribution in both the forward and backward directions.  As a future study, it also would be interesting to vary the manner in which the TPB film is prepared.  Additionally, our lamp intensity cuts off at approximately 110 nm because of the transmittance of its MgF$_{2}$ window (see Figure \ref{fig:ScintSpec}).  Clearly, it would be advantageous to extend these efficiency measurements down to 70--80 nm so that we could characterize these films near the scintillation wavelengths of neon and helium as well.  It is also possible that we could examine other fluor molecules as candidates for use in noble element scintillation detectors.  All of these efforts will help to continue to make these noble element scintillation detectors a viable technology in the coming years, and will continue to push forward the optimization of this radiation detection technique.

\section{Acknowledgments}\label{sec:Ack}
This work was supported by Los Alamos National Laboratory's Directed Research and Development program, the Department of Energy's NA-22 Office of Nonproliferation Research \& Development, National Science Foundation grant number PHY-0758120, and a Research Excellence Development grant from the University of South Dakota.  The authors thank Victor H. Gehman, Jr. and Stephen H. Jaditz for their careful reading and constructive comments.  The authors would also like to thank Hugh Lippincott, James Nikkel and Dan McKinsey for the preparation of the sample disks and valuable discussions regarding detection of EUV scintillation light with TPB as well as Andrew M. Dattelbaum and Anatoly V. Efimov for assistance with reflectometry measurements of our TPB film thickness.


\end{document}